# Maneuver-based Driving for Intervention in Autonomous Cars

**Henrik Detjen**
**Stefan Geisler**
henrik.detjen@hs-ruhrwest.de
stefan.geisler@hs-ruhrwest.de
University of Applied Sciences Ruhr West
Bottrop, DE

**Stefan Schneegass**
stefan.schneegass@uni-due.de
University of Duisburg-Essen
Essen, DE

## ABSTRACT

The way we communicate with autonomous cars will fundamentally change as soon as manual input is no longer required as back-up for the autonomous system. Maneuver-based driving is a potential way to allow still the user to intervene with the autonomous car to communicate requests such as stopping at the next parking lot. In this work, we highlight different research questions that still need to be explored to gain insights into how such control can be realized in the future.

## MANEUVER-BASED DRIVING AND INTERFACE PARADIGM CHANGE

Autonomous cars take over control from the driver. In SEA Level 5 [7], cars autonomously drive to the dedicated destination, and the driver can perform non-driving related tasks. These tasks include reading newspapers, working, or texting [5]. While, in theory, no additional input is possible (i.e., correcting the autonomous car), the passengers might still want to provide input from time to time. A reason for this could be, for example, overtaking a truck that blocks the scenic view or exiting the road in order to take a break and get a meal. In contrast to current input, the input necessary for such commands are on a maneuver level [4], not on a fine-grained control level (e.g., overtake or stop at the next parking lot instead of changing the driving direction to the left or accelerating manually). Kauer, Schreiber & Bruder [2] define a driver-centered maneuver command set. These kinds of interventions require a novel type of in-vehicle control since the traditional automotive user interfaces (i.e., steering wheel and pedal) are not designed to enter such complex maneuvers. Given that these maneuvers will only be done from time to time, the classical controls might not be the best fit, because drivers will not be trained as useful as they currently are in the future. Thus, we need to investigate new means of communicating with autonomous cars. In this work, we state different research questions that need to be investigated to allow maneuver-based driving to be a possible way of intervening with autonomous vehicles. We highlight these questions and offer potential design solutions.

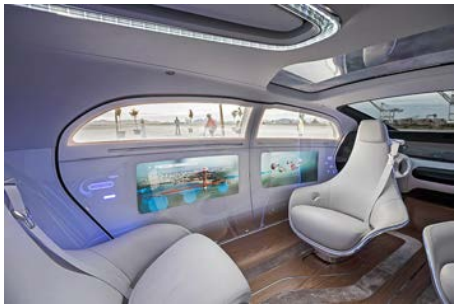

**Figure 1: Mecedes F 015 Concept Car Interior Design [3] - SAE Level 5**

### How to communicate available interventions?

Interventions will not be available in all situations. They require a particular environment (e.g., the car can only stop right if there is a parking space) and might also depend on the current speed of driving (e.g., the car needs a certain distance to decrease the speed to turn right). Thus, it is essential to communicate what options are available. One easy way of communicating these options is through augmented reality in the windshield [1]. The specific options are highlighted, and the driver can choose one of them.

### How to enter intervention commands?

Traditionally, buttons and knobs are used as a primary input. For autonomous cars, the user might be engaged with several different activities and might not even face forward (cf., Figure 1). Thus, different types of input commands are necessary that perhaps not even require physical controls at a specific part of the vehicle. Potential examples for such input include mid-air gestures or voice commands (e.g., [8] combine speech commands with pointing gestures for selection of a parking lot).

**How to provide feedback to the user?**

While the overall goal of the car might be clear to the passenger, communication of the current action of the car can still be of benefit to provide a feeling of control to the user [6]. Also, transparent communication of the current vehicle actions increases the driver's trust in the system [9]. However, it is still unclear what information should be communicated and how they should be communicated. When the next maneuver would be an overtaking of the autonomous car, it might be beneficial to highlight this to the user. One example would be a maneuver list that shows the upcoming maneuvers. Further, events that require the driver to intervene might also be communicated to the passenger. For example, the entered destination can no longer be reached (e.g., through a blocked road), and the autonomous vehicle needs further input on what alternative destination it should approach.

## CONCLUSION

Maneuver-based control of autonomous cars is promising in means of combining the strength of autonomous cars with the needs and requirements of passengers. While the vision of fully autonomous cars might render the driver obsolete, there are still situations in which an input of the driver is required. In this work, we highlight research questions that need to be tackled in the future to investigate whether maneuver-based driving can ease the challenges of future autonomous systems.